\documentclass[aps,prb,twocolumn,showpacs,superscriptaddress]{revtex4}

\begin{document}

\title{Comment on ``Conduction states in oxide perovskites: Three
manifestations of Ti$^{3 + }$ Jahn-Teller polarons in barium
titanate''}
\author{S. A. Prosandeev}
\affiliation{Physics Department, Rostov State University, 5 Zorge
St., 344090 Rostov on Don, Russia}
\author{, I. P. Raevski}
\affiliation{Physics Department, Rostov State University, 5 Zorge
St., 344090 Rostov on Don, Russia} \affiliation{Research Institute
of Physics, Rostov State University, 190 Stachki ave., 344090
Rostov on Don, Russia}
\author{M. A. Bunin}
\affiliation{Research Institute of Physics, Rostov State
University, 190 Stachki ave., 344090 Rostov on Don, Russia}

\maketitle

We have a remark to Sec. 6 were the authors \cite{Schirmer}
discuss polarons bound to doubly charged oxygen vacancies
$V_\mathrm{O}^{++}$.

By using EPR spectroscopy, Scharfschwerdt et. al. \cite{Scharf}
found in reduced BaTiO$_3$~ a signal which they attributed to a
symmetry broken state of Ti$^{3 + }$~ bound to
$V_\mathrm{O}^{++}$~ ($V_\mathrm{O}^{++}$Ti$^{3+}$). Later on,
theoretical studies performed in the framework of a Green function
approach considered possible reasons for such symmetry breaking.
\cite{Osipenko,JETP} Within that study the electron's activation
energies were obtained in accordance with data on
electroconductivity \cite{Raevski} which we will discuss below.

Theoretical studies \cite{Osipenko,JETP} did not consider the
states of the $F$-center-type (a vacancy centered, symmetric
electron density), but, later, computations showed that the states
of the $F$-center type , in the single charged oxygen vacancy
($V_\mathrm{O}^+$), can have lower energy than the states of the
$V_\mathrm{O}^{++}$Ti$^{3+}$-type \cite{Donnerberg,Bunin1,
Bunin2}. However, the EPR signal of such states has not been found
experimentally. \cite{Schirmer}

Hence, there is a seeming contradiction between the experiment
\cite{Scharf} and results of theoretical computations.
\cite{Donnerberg} In order to resolve this contradiction, Lenjer
et. al. \cite{Schirmer} suggest that the oxygen vacancy
$V_\mathrm{O}$~ in BaTiO$_3$~ is a negative-$U$~ center. This
could explain the absence of the EPR signal of $V_\mathrm{O}^+$.
The presence in experiment \cite{Scharf} of the EPR signal of a
symmetry broken state Lenjer et. al. \cite{Schirmer} relate to an
overpopulated state of the oxygen vacancy having 3 electrons
($V_\mathrm{O}$Ti$^{3+}$). The third electron, in the authors'
opinion,\cite{Schirmer} sits at the Ti site nearest to
$V_\mathrm{O}$ but the source of the existence of this third
electron on $V_\mathrm{O}$~ was not discussed, but, probably, it
can arise due to Nb doping of BaTiO$_3$~ which Scharfschwerdt et.
al. \cite{Scharf} employed in their experiment in order to
compensate acceptor-type Na impurities.

The main assumption of Lenjer et. al. \cite{Schirmer} regarding
the negative-$U$~ center seems to contradict the theoretical
computations \cite{Donnerberg,Eglitis} which obtained that
$V_\mathrm{O}$~ is a positive $U$-center. There can be also
contradictions with experimental data \cite{Raevski} on
electroconductivity which we want to discuss in more details.

It is well-known that the electroconductivity of perovskites
strongly depends on the degree of the reduction of the
sample.\cite{Raevski} The crystals in which the donors are nearly
fully compensated by acceptors are transparent and have low
conductivity. The strongly reduced samples are often dark (and
even black) and have a strong conductivity of n-type. Notice that,
during the reduction process, only the $V_\mathrm{O}$~
concentration is increased. Hence, the average number of electrons
on $V_\mathrm{O}$~ in this experiment cannot be more than 2.
However. in order that the room conductivity is high, the electron
energy levels of $V_\mathrm{O}$~ should be small, about 0.1
eV.\cite{Raevski}   The activation energy of $V_\mathrm{O}^+$~ we
will discuss below.

In disagreement with these experimental facts INDO computation
\cite{Eglitis} predicts that the $V_\mathrm{O}^+$~ electronic
energy level in KNbO$_3$~ is about 0.6 eV above the \emph{top of
the valence band} (2.7 eV below the bottom of the conduction
band), and the difference between the energies of $V_\mathrm{O}$~
and $V_\mathrm{O}^+$~ is positive, and it is about 0.3 eV. This is
a mystery, how it is possible to explain the high electric
conductivity and black color of the heavily reduced samples of
perovskites together with the high activation energy of electrons
in $V_\mathrm{O}$? However this computation is in line with other
computations of $V_\mathrm{O}$~ in perovskites showing the
presence of deep $V_\mathrm{O}$~ states of the $F$-center type.
\cite{Donnerberg,Park,Christensen,Bunin1,Bunin2} This puzzle
should be resolved somehow.

Experimental data on electroconductivity in SrTiO$_3$~ show that,
depending on the degree of reduction, the Fermi energy $E_F$~ in
the expression $\sigma \sim \exp\left[(E_F-E)/k_BT\right]$~
changes (here $E$~ is the energy of the bottom of the contactance
band): there were observed $E-E_F =$ 0.35 eV, 0.18 eV, and $\le$~
0.1 eV. The first value was related to the case when the
population of $V_\mathrm{O}^+$~ is very small and $E_F$~ coincides
with the electronic energy level of $V_\mathrm{O}^+$. The second
value (0.18 eV) corresponds to the case when the population of
$V_\mathrm{O}^+$~ is high and $E_F$~ lies in the middle between
the energy level of $V_\mathrm{O}^+$~ and the bottom of the
conduction band. The final, $\le$~ 0.1 eV, small energy was
related to the case of $V_\mathrm{O}$~ with 2 electrons. From
similar analysis for BaTiO$_3$~ the following Fermi energies were
obtained: 0.55 eV, 0.28 eV, 0.1 eV. In CaTiO$_3$: 0.15 eV, 0.08
eV, and $\le$ 0.1 eV. From these data it was deduced that, in
BaTiO$_{3}$, the activation energies of $V_\mathrm{O}^+$~ are
about $E_{1}= 0.55$ eV and the activation energy of
$V_\mathrm{O}$~ is $E_{2} \le 0.1$ eV; in SrTiO$_{3}$, $E_{1}=
0.35$ eV and $E_{2} <0.1$ eV; and in CaTiO$_{3}$, $E_{1}=0.15$ eV
and $E_{2}<0.1$ eV \cite{Raevski}. Notice that these energies were
observed in samples having different degrees of reduction and, in
particular, different color, black or gray (very small activation
energies) or light-yellow (comparatively deep levels). No other
donors besides V$_\mathrm{O}$~ were used (there are no Na and Nb
additions in these experiments in contrast to the experiment of
Scharfschwerdt et. al. \cite{Scharf}).

In semiempirical theoretical studies \cite{Osipenko,JETP,Raevski}
it was obtained that the symmetry broken electronic state
$V_O^{++}Ti^{3+}$~ has the energy about 0.2 eV with respect to
\emph{the conduction band bottom} plus the polaronic energy
connected with the interaction of the microscopic dipole with
lattice polarization. This is in very good agreement with the data
on electroconductivity. \cite{Raevski} However this scheme
contradicts the results of modern embedded cluster computations.
\cite{Donnerberg}

In principle, one could connect the first two values of the Fermi
energy (for instance, in BaTiO$_3$, 0.55 eV and 0.28 eV with
respect to the bottom of the conduction band) \cite{Raevski} with
$V_\mathrm{O}^+$~ and $V_\mathrm{O}$~ respectively. However, it is
not clear again the origin of the value $E-E_F \approx 0.1$~ eV as
well as the large gain of electroconductivity in heavily reduced
samples and their black color. Besides the explanation given
above, one could also consider $V_\mathrm{O}$~ clusters, surface
conductivity, and a strong decrease of the polaronic energy in
reduced samples. In this connection we want to cite reference
[\cite{Scott}] in which it was shown that the oxygen vacancies
have a tendency to ordering, especially when their concentration
increases. The $F$-center state can be destroyed in such pairs and
this can help appearing the low-energy electronic states.

Lenjer et al. \cite{Schirmer} also argue that, in experimental
study \cite{Mueller}, $V_\mathrm{O}^+$~ was not found. We stress
that $V_\mathrm{O}^+$~ exists in a thin interval of the relative
donor/acceptor concentration. In order to see these states one
should slowly vary the degree of the reduction or oxidation of the
sample. \cite{Raevski} At a given donor/acceptor concentration
ratio there are vacancies with presumably one charge state because
of a strong difference between the electronic energies of
$V_\mathrm{O}$~ and $V_\mathrm{O}^+$. \cite{Raevski}

The electroconductivity data \cite{Raevski} are consistent with
the existence of three possible states of V$_\mathrm{O}$~ having
zero, one or two electrons respectively (the first state having
zero electrons does not contribute to the conductivity). The small
value of one of the activation energies is connected with the
state of $V_\mathrm{O}$~ having two electrons: the electrostatic
interaction between these electrons makes this state rather
shallow. The comparatively large activation energy is connected
with $V_\mathrm{O}^+$. The large difference between these energies
in BaTiO$_{3}$~ and CaTiO$_{3}$~ was explained by stronger
polaronic effect in BaTiO$_{3}$~ due to softer lattice dynamics
\cite{Raevski}. We want also to cite a study in which it was shown
that reduced samples of barium titanate exhibit paramagnetic
susceptibility \cite{Kvantov}. All these experimental findings are
consistent with the assumption of the existence of
$V_\mathrm{O}^+$~ although its appearance requires rather strong
inequalities on the donor/acceptor concentration ratio
\cite{Raevski}. Other possible explanations of these data and new
schemes of computations should be explored in order to understand
the origin of the contradiction of this scheme with present time
computations.

Unfortunately, it is not clear from the experimental data on EPR
\cite{Schirmer} if the thermal activation energy of the ``third
electron on V$_\mathrm{O}$'' is comparable with the activation
energy of ``V$_\mathrm{O}^+$'' obtained from electroconductivity.
\cite{Raevski} In any case, it would be helpful to analyze the EPR
data together with data on electroconductivity and, perhaps, also
on optics (see discussions, for instance, in
\cite{Bursian,Raevski}). In our opinion, first-principles
computations should be also developed further in order to explain
the experimentally observed activation energies connected with
$V_\mathrm{O}$~ and their tendencies in sequences of perovskites.
\cite{Raevski} Finally, the electronic structure of
$V_\mathrm{O}$~ and $V_\mathrm{O}^+$~ remain a mystery and further
studies are necessary.

This study is partially supported by RFBR grants 01-02-16029 and
01-03-33119.

\end{document}